\documentclass[showpacs,floatfix,superscriptaddress,showpacs,twocolumn,amssymb,amsfonts,prb,aps]{revtex4}% Physical Review Letters
\usepackage{longtable,graphicx,epsfig,dcolumn}

\begin{document}
\bibliographystyle{revtex}

\title{
Wetting Phase Transition at the Surface of Liquid Ga-Bi alloys:\\
An X-ray Reflectivity Study}

\author{H.~Tostmann}
\affiliation{Division of Applied Sciences and Department of
Physics, Harvard University, Cambridge MA 02138}

\author{E.~DiMasi}
\affiliation{Department of Physics, Brookhaven National
Laboratory, Upton NY 11973-5000}

\author{P.~S.~Pershan}
\affiliation{Division of Applied Sciences and Department of
Physics, Harvard University, Cambridge MA 02138}

\author{B.~M.~Ocko}
\affiliation{Department of Physics, Brookhaven National
Laboratory, Upton NY 11973-5000}

\author{O.~G.~Shpyrko}
\affiliation{Division of Applied Sciences and Department of
Physics, Harvard University, Cambridge MA 02138}

\author{M.~Deutsch}
\affiliation{Department of Physics, Bar-Ilan University, Ramat-Gan
52100, Israel}
\date{Dated: 12 July 1999}

\begin{abstract}
 X-ray reflectivity measurements of the  binary liquid Ga-Bi  alloy
 reveal a dramatically different surface structure
 above and below the monotectic temperature $T_{mono}=222^{\circ}$\,C.
 A Gibbs-adsorbed Bi monolayer resides at the surface
 at both regimes. However,
 a 30\,{\AA} thick, Bi-rich wetting film intrudes
 between the Bi monolayer and the Ga-rich bulk for $T > T_{mono}$.
 The  internal structure of the wetting film
 is determined with {\AA} resolution,
 showing a theoretically unexpected concentration gradient
 and a highly diffuse interface with the bulk phase.

\end{abstract}


% 61.25.Mv  liquid metals and liquid alloys
% 68.10.--m fluid surfaces and fluid--fluid interfaces
% 61.10.--i x-ray determination of structures
\maketitle

\section{Introduction}
A thick wetting film may be stable at the free surface of a binary
immiscible liquid mixture and the temperature dependent formation of this
surface film is strongly influenced by the bulk critical demixing
of the underlying bulk phase.\cite{cahn}
In contrast to surface segregation, where a monolayer of
the low surface tension component forms at the interface of the binary mixture,
wetting is a genuine phase transition occuring at
the interface resulting in the formation of a  mesoscopically
or macroscopically thick film.\cite{dietrich}
To date, nearly all studies of wetting phenomena have been carried out with
dielectric liquids dominated by long-range van der Waals interactions\cite{domb} and the
experimental techniques used achieve mesoscopic resolution only.
For binary alloys dominated by screened Coulomb interactions,
evidence for a wetting transition has
so far  been obtained  only for  Ga-Bi\cite{nattland} and Ga-Pb.\cite{chatain}
In the case of Ga-Bi,  the formation of a Bi-rich wetting film has been detected by
ellipsometry in the temperature range from 220$^{\circ}$\,C to
228$^{\circ}$\,C.\cite{nattland}
However, since the film thickness is much smaller than the wavelength of visible light, ellipsometry
cannot provide {\AA}-resolution structural information.
By contrast, x-ray surface scattering techniques allow  determination  of
the structure of the wetting film with atomic resolution
and address issues such as the internal structure of the
film and its  evolution from molecular to mesoscopic thickness, none of which
was hitherto addressed by any of the previous measurements.

\begin{figure}[tbp]
\epsfig{file=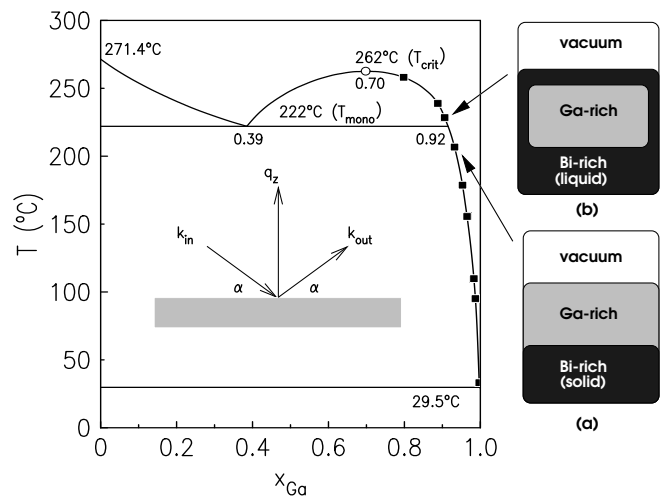,
angle=0, width=1.0\columnwidth} \label{fig:fig1} \caption{ Phase
diagram of Ga-Bi (Ref.\,8). Between 29.5$^{\circ}$\,C and
222$^{\circ}$\,C, solid Bi is in equilibrium with a Ga-rich liquid
phase. Between 222$^{\circ}$\,C and 262$^{\circ}$\,C, a liquid
Bi-rich phase is in coexistence with a liquid Ga-rich phase. The
surface structure of this alloy has been measured between
35$^{\circ}$\,C and 258$^{\circ}$\,C  at selected points along the
coexistence line (\protect\rule{1.5mm}{1.5mm}). For
$T<$222$^{\circ}$\,C   partial wetting occurs (a), whereas
complete wetting is found  for $T>$222$^{\circ}$\,C (b). The inset
depicts the geometry for specular x-ray reflectivity. }
\end{figure}

The wetting phase transition is depicted schematically in Fig.\,1.
Below the bulk critical point of demixing, $T_{crit}$, the bulk
phase separates into two immiscible phases, the high density
Bi-rich phase and the low density Ga-rich phase. Below the
characteristic wetting temperature $T_w < T_{crit}$, the high
density phase is confined to the bottom of the container as
expected (see Fig.~\ref{fig:fig1}(a)). As discussed in more detail
below, for Ga-Bi, the high density phase is solid in this case and
$T_w$ coincides with the monotectic temperature $T_{mono}$. In
contrast, above $T_w$ the high density phase completely wets the
free surface by intruding between the low density phase and the
gas phase in defiance of gravity (see
Fig.~\ref{fig:fig1}(b)).\cite{dietrich,rowlinson}

\section{Sample Preparation}
The Ga-Bi alloy was prepared
in a inert-gas box  using at least 99.9999\% pure metals.
A solid Bi ingot was placed in a Mo pan and supercooled liquid Ga
was added to cover the Bi ingot.
At room temperature,  the solubility of Bi in Ga is less than 0.2at\%.
Increasing the temperature results in  continuously dissolving  more Bi in the Ga-rich phase with
 solid Bi remaining at the bottom of the pan  up to the monotectic
temperature, $T_{mono}$= 222$^{\circ}$\,C. The initial Bi content
was chosen to be high enough  that the two phase equilibrium along
the miscibility gap could be followed up to few degrees below
$T_{crit}$ without crossing into the homogeneous phase region (see
Fig.~\ref{fig:fig1}). The  temperature on the sample surface was
measured with a Mo coated thermocouple. The alloy was contained in
an  ultra high vacuum chamber and the residual oxide on the sample
was removed by sputtering with $\rm Ar^+$ ions.\cite{indium}
Surface sensitive x-ray reflectivity (XR) experiments were
carried out using the liquid surface spectrometer at  beamline
X22B  at the National Synchrotron Light Source with an  x-ray
wavelength $\lambda=$1.24\,{\AA} and a detector resolution of
$\Delta q_z=$0.03\,$\rm {\AA}^{-1}$. The background intensity, due
mainly to scattering from the bulk liquid, was subtracted from the
specular signal by displacing the detector out of the reflection
plane.\cite{indium}

\begin{figure}[tbp]\epsfig{file=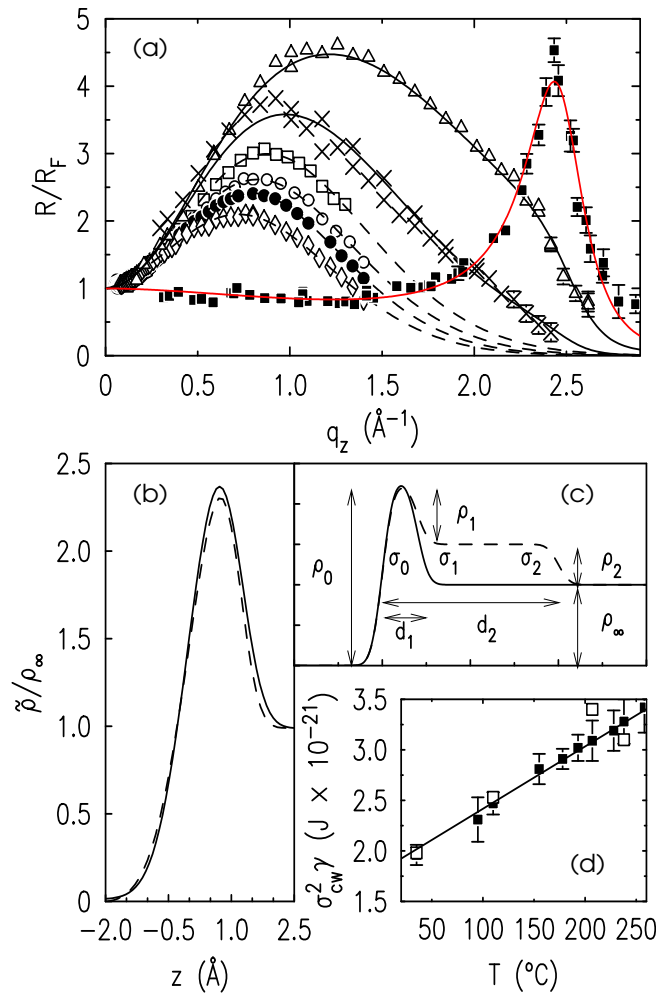,
angle=0, width=1.0\columnwidth} \label{fig:fig2}

\caption{
(a) Fresnel normalized reflectivity \protect$R/R_F$ from  Ga-Bi alloys
for \protect$T < T_{mono}$:
(\protect\rule{1.5mm}{1.5mm}): 100\,at\% Ga;
 (\protect$\triangle$): 99.7\,at\% Ga
(35\protect$^{\circ}$\,C); (\protect$\times$): 98.7\% Ga (95$^{\circ}$\,C);
(\protect$\Box$): 98.4\% Ga (110$^{\circ}$\,C );
(\protect$\circ$): 96.5\% Ga (155$^{\circ}$\,C );
(\protect$\bullet$): 95.3\% Ga (178$^{\circ}$\,C ) and
(\protect$\diamond$): 93.2\% Ga (207$^{\circ}$\,C ).
Shaded line: fit to layered density profile of pure Ga (see Ref.\,9);
solid lines: fit to layered density profile of Ga plus Bi monolayer (see Ref.\,10);
broken lines: fit to Bi monolayer (this work).
(b) Intrinsic   density profiles
normalized to \protect$\rho_{\infty}$
for Ga-Bi for 110$^{\circ}$\,C $\leq T \leq$ 207$^{\circ}$\,C
using the one-box model described in the text. The profiles essentially
fall on top of each other and two representative fits are shown:
110$^{\circ}$\,C (solid line)
and 207$^{\circ}$\,C (broken line).
(c): Schematic representation  of the (two-)box model.
(d): CW roughness $\sigma_{cw}$ for Ga-Bi alloys as
a function of temperature using  $\gamma$ from
our fit (\protect\rule{1.5mm}{1.5mm}) and from macroscopic
surface tension measurements (\protect$\Box$, see Ref.\,14).
}
\end{figure}

\section{Experimental Techniques}
The  intensity reflected from the surface, $R(q_z)$, is measured
as a function of the normal component $q_z =
(4\pi/\lambda)\sin\alpha$ of the momentum transfer. The XR
geometry is depicted schematically in the inset of
Fig.~\ref{fig:fig1}. $R (q_z)$ therefore yields information about
the surface-normal structure as given by
\begin{equation}
    R(q_z) =  R_F(q_z)
        \left| \Phi (q_z) \right| ^2
    \exp [ - \sigma _{\mbox{\scriptsize cw}}^2q_z^2 ],
\label{eq:circle}
\end{equation}
where $R_F (q_z)$ is the Fresnel reflectivity of a flat, infinitely sharp  surface and $\Phi (q_z)$
is the Fourier transform of the local surface-normal  density profile $\tilde{\rho} (z)$:\cite{indium}
\begin{equation}
\Phi (q_z) = \frac{1}{\rho_{\infty}} \int dz \frac{d\tilde{\rho} (z) }{dz}
\exp(\imath q_z z)
\label{eq:structure}
\end{equation}
with the bulk electron density, $\rho_{\infty}$.
The exponential factor in Eq.\,\ref{eq:circle} accounts for roughening
of the intrinsic density profile $\tilde{\rho} (z)$ by  capillary waves (CW):
\begin{equation}
\sigma _{\mbox{\scriptsize cw}}^2 = \frac{k_B T}{2\pi \gamma }
    \ln \left(
    \frac{q_{\mbox{\scriptsize max}}}{q_{\mbox{\scriptsize res}}
    }\right)  .
\label{eq:cw}
\end{equation}
The CW spectrum  is cut off at
small $q_z$ by the detector resolution  $q_{res}$ and at large $q_z$ by the atomic size
$a$ with $q_{max} \approx \pi/a$.\cite{indium}
However, $\tilde{\rho} (z)$, can not be obtained directly from Eq.\,\ref{eq:structure}
and we resort to the widely accepted
procedure of adopting a physically motivated model for $\tilde{\rho}(z)$ and fitting its
Fourier transform  to the experimentally
determined $R (q_z)$ as will be shown further below.\cite{indium}

The reflectivity of pure liquid Ga exhibits a pronounced
interference peak indicating surface-induced layering of ions near
the surface (see Fig.~\ref{fig:fig2}(a)).\cite{gallium} Previous
experiments on Ga-Bi at low temperatures\cite{bunsen,rice} show
that the layering peak is suppressed in the liquid alloy and $R
(q_z)$ is dominated by a broad maximum consistent with a single Bi
monolayer segregated at the surface as expected since $\gamma$ is
considerably lower for Bi than for Ga.\cite{rice}

\section{Results}
Here,  we report x-ray measurements from liquid Ga-Bi  for $T$ up
to 258$^{\circ}$\,C encompassing the formation of the wetting film
at $T >T_{mono}$, the lowest temperature at which  the liquid
Bi-rich phase is stable. The actual wetting transition likely lies
below $T_{mono}$ but does not manifest itself since the wetting
film is solid for $T < T_{mono}$. Experimental evidence for this
assumption has been found for the similar systems
Ga-Pb\cite{wynblatt} and K-KCl.\cite{kkcl} The  discussion of the
data is divided into two parts: (i) $T < T_w \simeq T_{mono}$
(110$^{\circ}$\,C-207$^{\circ}$\,C) and (ii) $T > T_w$
(228$^{\circ}$\,C-258$^{\circ}$\,C). We present the results for $T
< T_{mono}$ (scenario (a) in Fig.~\ref{fig:fig1}) first since this
corresponds to the simpler surface structure with monolayer
segregation but no wetting film.

{\bf  Below $T_{mono}$:} Fig.~\ref{fig:fig2}(a) shows  $R (q_z)$
from liquid Ga-Bi at selected points along the coexistence line
between 110$^{\circ}$\,C  and 207$^{\circ}$\,C. Since subsurface
layering is impossible to resolve at these temperatures, we simply
model the near-surface density, which is dominated by the Bi
monolayer segregation, by one box of density $\rho_1$ and width
$d_1$. The density profiles are shown in Fig.~\ref{fig:fig2}(b)
and the box model is shown schematically in Fig.~\ref{fig:fig2}(c)
(solid line). The best fit to $R(q_z)$ that corresponds to this
density profile is represented by the broken lines in
Fig.~\ref{fig:fig2}(a). The mathematical description of the
general box model  with a maximum number  of two boxes is: {\small
\begin{equation}
\tilde{\rho}(z) = \frac{\rho_0}{2}
\left\{ 1 + {\rm erf} \left( \frac{z}{\sigma_{0}} \right) \right\}
 - \sum_{i=1}^2
\frac{\rho_i}{2} \left\{ 1 + {\rm erf} \left(\frac{z-d_i}{\sigma_{i}}\right)\right\}
\label{eq:box}
\end{equation}
} where $\sigma_0$ and $\sigma_{i}$ are the intrinsic
 roughnesses of each of the three interfaces.
The total roughness, $\sigma$, of the  interface between the vapor
and the outermost surface layer is  given by $\sigma^2 =
\sigma^2_{0} + \sigma^2_{cw}$ with $\sigma_0$ = 0.78$\pm
0.15$\,{\AA} for $T < T_{mono}$. As predicted by capillary wave
theory, the product $\sigma_{cw}^2 \gamma$ (with $\gamma$ from our
fits) depends linearly on $T$ (see Eq.\,\ref{eq:cw} and
(\protect\rule{1.5mm}{1.5mm}) in Fig.~\ref{fig:fig2}(d)) over the
entire temperature range.
 This variation is essentially the same if
$\gamma$ is not taken from our fits but from macroscopic surface
tension measurements (\protect$\Box$).\cite{tschirner} Apart from
the increasing surface roughness, the structure of the surface
does not change over the entire temperature range from
110$^{\circ}$\,C to 207$^{\circ}$\,C, as witnessed by the
intrinsic density profiles that fall right on top of each other
once the  theoretically predicted temperature dependence of the CW
roughness has been corrected for (Fig.~\ref{fig:fig2}(b)). Between
212$^{\circ}$\,C and 224$^{\circ}$\,C, we observed rapid and
random changes in the scattered intensity  over the entire $q_z$
range. This unstable behavior of the surface is most likely due to
the coexistence of patches of different film thickness induced by
the temperature gradient of about 6K normal to the  the sample
pan.

\begin{figure}[tbp]\epsfig{file=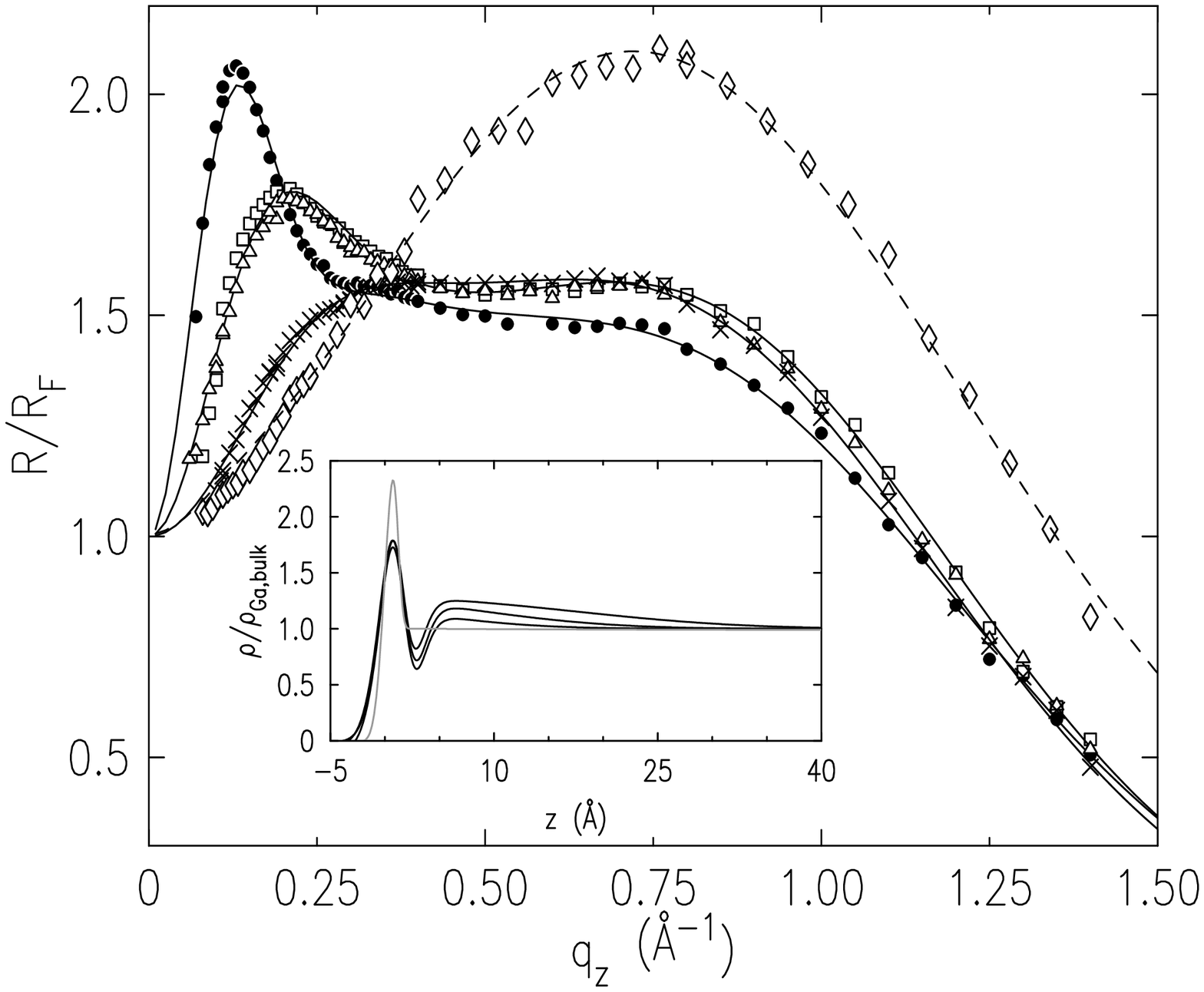,
angle=0, width=1.0\columnwidth} \label{fig:fig3}

\caption{ $R/R_F$ from Ga-Bi alloys for $T > T_w$: ($\bullet$):
90.6\,at\% Ga (228$^{\circ}$\,C); ($\Box$): 88.8\% Ga
(238$^{\circ}$\,C); ($\triangle$): ditto, but after heating to 258
$^{\circ}$\,C and cooling down; ($\times$): 79.1\% Ga
(258$^{\circ}$\,C); (for comparison, see:($\diamond$),
207$^{\circ}$\,C). Solid lines fit to two-box model; broken line:
fit to one-box model. Inset: intrinsic density profiles,
$\tilde{\rho}(z)$, for Ga-Bi alloys normalized to the bulk
density, $\rho_{Ga,bulk} = \rho_{\infty}$, at $T > T_w$: solid
lines ordered  in decreasing wetting film density:
228$^{\circ}$\,C,  238$^{\circ}$\,C and 258$^{\circ}$\,C. Compare
to the  one-box density profile of Ga-Bi at 207$^{\circ}$\,C
(shaded line). }
\end{figure}

{\bf Above $T_{mono}$:} The surface stabilized  above  $T_{mono}$
at about 228$^{\circ}$\,C and and a sharper peak in $R(q_z)$
appeared centered around 0.13\,$\rm {\AA}^{-1}$ (see
Fig.~\ref{fig:fig3}), indicating a thick film of high density
forming near the surface. The persistence of the  broad maximum
centered around $q_z \approx 0.75$\,{\AA}$^{-1}$ indicates that Bi
monolayer segregation is still present along with the newly formed
thick wetting film. Several models were used to fit $R\,(q_z)$ but
they all result in essentially the same density profiles. Here, we
use the simple two-box model (see broken line in
Fig.~\ref{fig:fig2}(c) and Eq.\,\ref{eq:box}). As can be seen in
Fig.~\ref{fig:fig3}, this simple model gives an excellent
description of the experimentally obtained reflectivity. The
pertinent density profiles describing the surface-normal structure
of Ga-Bi alloys for $T > T_{mono}$  are shown in the inset of
Fig.~\ref{fig:fig3} and  have the following  features that can be
compared to present predictions of wetting theory: (i) density,
(ii) thickness and (iii) roughness of the wetting film.

(i) The concentration profile is highly inhomogeneous with a high
density region at the outermost surface layer.\cite{dip} The
integrated density of this adlayer is consistent with a monolayer
of the same density as the monolayer found for $T < T_{mono}$.
However, the Bi monolayer segregated on top of the wetting film is
rougher and broader than the Bi monolayer segregated on top of the
Ga-rich bulk phase for $T < T_{mono}$. This change in surface
structure corresponds to a change in the intrinsic roughness of
the interface liquid/vacuum, $\sigma_0$, from 0.78$\pm$0.15\,{\AA}
below $T_{mono}$  to 1.7$\pm$0.3\,{\AA} above $T_{mono}$. However,
the CW roughness still follows the predicted T-dependence of
Eq.\,3 (see Fig.~\ref{fig:fig2}(d)).

This change of $\sigma_0$ upon crossing $T_{mono}$
is possibly related to the fact that for $T > T_{mono}$,
the pure Bi monolayer segregates against a Bi-rich wetting film made out of atoms of
different size and with repulsive heteroatomic interactions. By contrast,
for $T < T_{mono}$,
this segregation takes place against
almost pure Ga.
The fact that the Bi monolayer segregation persists from partial wetting to complete wetting
means that the wetting film does not intrude between the Ga-rich bulk phase and vacuum,
as assumed by wetting theory, but between the Bi monolayer already segregated at the surface
and the Ga-rich bulk phase. This should have a pronounced influence on the energy balance
at the surface which ultimately governs the wetting phase transition. Even though the
possibility of an unspecified  concentration gradient is included in Cahn's general theory\cite{cahn},
to our knowledge, an
inhomogeneous density profile is neither treated explicitely in theoretical calculations
(see the homogeneous profiles displayed in\,\cite{profile,beysens}) nor found experimentally.
 At 228$^{\circ}$\,C the density of the thick wetting film is $\rho_2 =(1.25\pm 0.03)\rho_{\infty}$
where $\rho_{\infty}$ is the  density of the Ga-rich bulk phase. This agrees very well with the ratio of
the densities of the coexisting  Bi-rich and  Ga-rich phases in the bulk, calculated from the phase diagram
to be 1.23. The density of the wetting film reaches $\rho_{\infty}$
 with increasing T ($\rho_2/\rho_{\infty}=$ 1.18$\pm 0.03$ at
238$^{\circ}$\,C and 1.08$\pm 0.02$ at 258$^{\circ}$\,C)
as the  densities of the two  bulk phases converge upon
 approaching $T_{crit}$  (their density  ratio is 1.20 at 238$^{\circ}$\,C and
1.08 at 258$^{\circ}$\,C).
This  strongly supports the conclusion that the thick wetting film
is the Bi-rich bulk phase as predicted by wetting theory.

(ii)  The thickness of the wetting film, $d_2$, at
228$^{\circ}$\,C  is determined from the density profile to be
$\sim$30\,{\AA}, consistent with a model-dependent estimate from
ellipsometry results.\cite{nattland} In addition, the thickness of
the  wetting film is corroborated by independent grazing incidence
diffraction experiments resolving the in-plane structure which
will be reported elsewhere.\cite{iop} The wetting film in this
Coulomb liquid with short-range interactions is considerably
thinner than wetting films that have been observed in dielectric
liquids with long-range interactions.\cite{domb,kayser} An
important question concerning the thickness of a wetting film is
whether the wetting film has been investigated in equilibrium. The
fact that identical reflectivities were measured at
238$^{\circ}$\,C taken 24 hours apart (Fig.~\ref{fig:fig3}), is
strong evidence that the film thickness is in equilibrium in our
study.\cite{kayser}

(iii) The roughness between the Bi-rich wetting film and the
Ga-rich bulk film is much higher that the roughness of the free
interface LM/vacuum. This is be expected since the interfacial
tension between two similar liquids is generally much lower than
the liquid/gas surface tension.\cite{israel} On the other hand, it
is not clear how sharp a concentration gradient should be expected
between the Ga-rich and the Bi-rich phase and the interface may be
essentially diffuse, independent of CW roughness. With increasing
temperature, the wetting film becomes less well defined and it is
not obvious whether this is due to the fact that the film is
becoming slightly thinner, or if only the interface between two
converging phases gets more diffuse. At any rate, we are far
enough away from $T_{crit}$ that the predicted thickening of the
wetting film  due to the increasing correlation length of the
concentration fluctuations should not play a role. Once the thick
wetting film has formed, it is not possible to cool the sample
below $T_{mono}$ since the Bi-rich wetting film remains at the
surface and freezes.

\section{Summary}
In summary, we investigated the structural changes occurring
on  atomic length scales at the  surface
of  liquid Ga-Bi during the wetting transition.
In the case of partial wetting ($T < T_{mono}$),
a Bi monolayer segregates at the surface to lower the surface energy. Above
the monotectic temperature complete wetting is found and
a  30\,{\AA} thick wetting film intrudes between this monolayer and the Ga-rich
bulk phase. This is the first time that the microscopic structure of a wetting film has been
studied and  that the concentration profile of the wetting film has
been shown to vary on a level of several atomic diameters.

\section{Acknowledgments}
This work is supported by the U.S.~DOE
Grant No. DE-FG02-88-ER45379.
%and the U.S.--Israel
%Binational Science Foundation, Jerusalem.
Brookhaven National Laboratory is supported by
U.S.\ DOE Contract No.\ DE-AC02-98CH10886.
%HT acknowledges support from the Deutsche Forschungsgemeinschaft.

%   %   %   references  %   %   %

\bibliographystyle{unsrt}

\begin{thebibliography}{10}
\bibitem{dip} The dip in density between the outermost Bi adlayer and the thick wetting layer
can be modeled with an atomic   model similar to the one used for Bi segregated on top of Ga
with subsurface layering (see Ref.\,10).
Including subsurface layering  provides a slightly improved
fit but does not substantially alter the parameters of the wetting layer.
\end{thebibliography}

\end{document}